\documentclass{pasj00}
\draft

\begin{document}
\SetRunningHead{T. Hamada, T. Fukushige, and J. Makino}{PGPG}
\Received{2004/07/27}
\Accepted{2005/07/20}

\title{PGPG: An Automatic Generator of Pipeline Design for Programmable GRAPE Systems}

\author{Tsuyoshi \textsc{Hamada} and Toshiyuki \textsc{Fukushige}}
\affil{Department of General System Studies, College of Arts and Sciences,\\
University of Tokyo, 3-8-1 Komaba, Meguro, Tokyo 153-8902}
\email{hamada@provence.c.u-tokyo.ac.jp, fukushig@provence.c.u-tokyo.ac.jp}

\and
\author{Junichiro \textsc{Makino}}%
\affil{Department of Astronomy, School of Science, \\
University of Tokyo, 7-3-1 Hongo, Bunkyo, Tokyo 133-0033}
\email{makino@astron.s.u-tokyo.ac.jp}


%


\maketitle

\newpage

\begin{abstract}

We have developed PGPG (Pipeline Generator for Programmable GRAPE), a
software which generates the low-level design of the pipeline processor
and communication software for FPGA-based computing engines (FBCEs).
An FBCE typically consists of one or multiple FPGA (Field-Programmable
Gate Array) chips and local memory. Here, the term
``Field-Programmable'' means that one can rewrite the logic
implemented to the chip after the hardware is completed, and therefore
a single FBCE can be used for calculation of various functions, for
example pipeline processors for gravity, SPH interaction, or image
processing. The main problem with FBCEs is that the user need to
develop the detailed hardware design for the processor to be
implemented to FPGA chips. In addition, she or he has to write the
control logic for the processor, communication and data conversion
library on the host processor, and application program which uses the
developed processor. These require detailed knowledge of hardware
design, a hardware description language such as VHDL, the operating
system and the application, and amount of human work is huge. A
relatively simple design would require 1 person-year or more.  The
PGPG software generates all necessary design descriptions, except for
the application software itself, from a high-level design description
of the pipeline processor in the PGPG language. The PGPG language is a
simple language, specialized to the description of pipeline
processors. Thus, the design of pipeline processor in PGPG language is
much easier than the traditional design. For real applications such as
the pipeline for gravitational interaction, the pipeline processor
generated by PGPG achieved the performance similar to that of
hand-written code. In this paper we present a detailed description of
PGPG version 1.0.

\end{abstract}

\section{Introduction}
\label{secintro}

Astronomical many-body simulations have been widely used to
investigate the formation and evolution of various astronomical
systems, such as planetary systems, globular clusters, galaxies,
clusters of galaxies, and large scale structures.  In such
simulations, we treat planetesimals, stars, or galaxies as particles
interacting with each other. We numerically evaluate interactions
between the particles and advance the particles according to Newton's
equation of motion.

In many cases, the size of an astrophysical many-body simulation is
limited by the available computational resources.  Simulation of pure
gravitational many-body system is a typical example.  Since the gravity
is a long-range interaction, the calculation cost is $O(N^2)$ per
timestep for the simplest scheme, where $N$ is the number of particles
in the system.  We can reduce this $O(N^2)$ calculation cost to $O(N\log
N)$, by using some approximated algorithms, such as the Barnes-Hut
treecode (Barnes, Hut 1986), but the scaling coefficient is pretty
large.  Thus, the calculation of the interaction between particles is
usually the most expensive part of the entire calculation, and thus
limits the number of particles we can handle.  Smoothed Particle
Hydrodynamics (SPH, Lucy 1977, Gingold, Monaghan 1977), in which
particles are used to represent the fluid, is another example.  In SPH
calculations, hydrodynamical equation is expressed by short-range
interaction between particles.  The calculation cost of this SPH
interaction is rather high, because the average number of particles
which interact with one particle is fairly large, typically around 50,
and the calculation of single pairwise interaction is quite a bit more
complex compared to gravitational interaction. 

Astrophysics is not the only field where the particle-based calculation
is used.  Molecular dynamics (MD) simulation and boundary element method
(BEM) are examples of numerical methods where each element of the system
in principle interacts with all other elements in the system.  In both
cases, approaches similar to Barnes-Hut treecode or FMM (Greengard,
Rokhlin 1987) help to reduce the calculation cost, but the interaction
calculation dominates the total calculation cost. 

One extreme approach to accelerate the particle-based simulation is to
build a special-purpose computer for the interaction calculation.  Two
characteristics of the interaction calculation make it well suited for
such approach.  Firstly, the calculation of pairwise interaction is
relatively simple.  In the case of gravitational interaction, the total
number of floating-point operations (counting all operations, including
square root and divide operations) is only 20.  So it is not
inconceivable to design a fully pipelined, hardwired processor dedicated
to the calculation of gravitational interaction.  For other application
like SPH or molecular dynamics, the interaction calculation is more
complicated, but still hardware approach is feasible.  Secondly, the
interaction is in its simplest form all-to-all.  In other words, each
particle interacts with all other particles in the system.  Thus, there
is lots of parallelism available.  In particular, it is possible to
design a hardware so that it calculate the force from one particle to
many other particles in parallel.  In this way we can reduce the
required memory bandwidth.  Of course, if the interaction is of
short-range nature, one need to implement some clever way to reduce
calculation cost from $O(N^2)$ to $O(N)$, and the reduction in the
memory bandwidth is not as effective as in the case of true $O(N^2)$
calculation. 

The approach to develop specialized hardware for gravitational
interaction, materialized in the GRAPE ("GRAvity piPE") project
(Sugimoto et al. 1990; Makino and Taiji 1998), has been fairly
successful, achieving the speed comparable or faster than the fastest
general-purpose computers for the price tag one or two orders of
magnitude smaller. For example, GRAPE-6, which costed 500M JYE,
achieved the peak speed of 64 Tflops. This speed is favorably compared
to the peak speed of the Earth Simulator (40Tflops) or
ASCI-Q(30Tflops), both costed several tens of billions of JYE. A major
limitation of GRAPE is that it cannot handle anything other than the
interaction through $1/r$ potential. It is certainly possible to build
a hardware that can handle arbitrary central force, so that molecular
dynamics calculation can also be handled (Ito et al. 1993; Fukushige et al
1996; Narumi et al. 1999; Taiji et al 2003). 

However, to design a hardware that can calculate both the
gravitational interaction and, for example, an SPH interaction is
quite difficult. Actually, to develop the pipeline processor just for
SPH interaction turned out to be a rather difficult task (Yokono et
al. 1999). This is provably because the SPH interaction is much more
complex than gravity.

Computing devices which uses FPGA (Field-Programmable Gate Array)
chips could offer the level of flexibility that was impossible to
achieve with the conventional GRAPE approaches. As its name suggests,
FPGA is a mass-produced LSI chip, consisting of a large number of
logic elements and switching network. By programming these logic
elements and switching network, we can implement an arbitrary logic
design, as far as it can fit to the chip used. Thus, a single hardware
can be used to implement various pipeline processors, such as that for
gravity, SPH, and others. Such FPGA-based ``reconfigurable'' computing
device has been an active area of research since Splash-1 and Splash-2
(Buell et al. 1996), and several groups, including ourselves, have
tried to apply the idea of reconfigurable computing to particle
simulations (Kim et al. 1995, Hamada et al. 2000, Spurzem et
al. 2002). Hamada et al. (2000) called this approach ``Programmable
GRAPE'' or PROGRAPE.

\begin{figure}
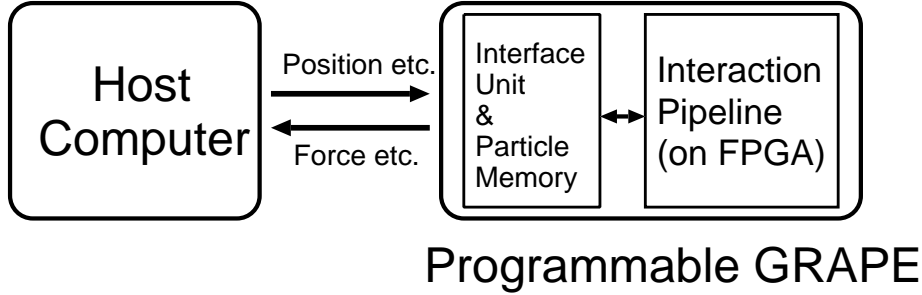

\begin{center}
\leavevmode
\FigureFile(12 cm,8 cm){./figure1.eps}
\end{center}
\caption{Basic structure of a programmable GRAPE(PROGRAPE).}
\label{fig1}
\end{figure}

Figure \ref{fig1} shows the  basic structure of a PROGRAPE system.
It consists
of a programmable GRAPE hardware and a host computer. The programmable
GRAPE hardware typically is composed of FPGA chips to which the
interaction pipelines are implemented, a particle memory, and an
interface unit, and calculates the
interaction $\mathbf{ f}_i$ between $i$-the particle and other particles
expressed as
\begin{equation}
\mathbf{ f}_{i} = \sum_{j}\mathbf{ G}(\mathbf{ a}_{i},\mathbf{
a}_{j}),
\label{eq:PGPGbasic}
\end{equation}
where $\mathbf{ a}_{i}$ is the physical value of the $i$-th particle,
such as position and velocities, and $\mathbf{ G}()$ is a user-specified
function. We specify the function $\mathbf{ G}()$ by programming
FPGA. The physical values $\mathbf{ a}_{j}$ of all particles are stored
in the particle memory and supplies them to the interaction
pipeline. The physical values $\mathbf{ a}_i$ are stored in registers of
the interaction pipeline.  The interface unit controls communications
between the programmable GRAPE hardware and the host computer. The
host computer performs all other calculations.

FPGA-based PROGRAPEs have several important advantages over
conventional full-custom GRAPE processors. One is that the development
cost of the chip itself is paid by the manufacturer of the chip, not
by us. Thus, initial cost is much lower. This low development cost
means that new hardwares can be developed in shorter cycle. Large
GRAPE hardwares took several years to develop, and this means the
device technology used in GRAPE hardwares, even at the time of its
completion, is a few years old. This delay implies quite a large
performance hit.

Thus, even though the efficiency in the transistor usage is much worse
than full-custom GRAPE processors, the actual price-performance of a
PROGRAPE system is not so bad, if one condition is satisfied: If the
design of the pipeline processor to be implemented in FPGA and other
necessary softwares can be developed sufficiently fast. Previous
experiences tell us that it is not the case. To implement a relatively
simple pipeline for gravitational interaction calculation took more
than one person-year, and implementation of even a simple SPH pipeline
would take much more. Thus, clearly the difficulty of the software
development has been the limiting factor for the practical use of
PROGRAPE or other FPGA-based computing device.

The difficulty is partly because we have
to design the interaction pipeline itself, for which we need rather
detailed and lengthy description of hardware logic in
hardware-description languages such as VHDL. In addition to the
pipeline itself, we also need to develop the control logic for the
pipeline and communication to the host, driver software on the host
computer, and software emulator library used to verify the design
(see section \ref{sectrad}).

In theory, most of the design description of softwares and hardwares,
including the bit-level design of the interaction pipeline itself, can
be automatically generated from some high-level description of the
pipeline itself. The basic idea behind the PGPG (Pipeline Generator
for Programmable GRAPE) system, which we describe in this paper, is to
realize such automatic generation. PGPG generates all necessary
hardware design descriptions and driver softwares, from high-level
description of the pipeline processor itself. Thus, the user is
relieved of the burden of learning complex VHDL language.
Also, the driver software is automatically generated, so
that the user can concentrate on writing the application program, not
the low-level driver software for a specific hardware. Thus, we can
dramatically reduce the amount of the work of the application
programmer. More importantly, when a new hardware becomes ready, once
the PGPG system is ported, all user applications developed on it
works unchanged. The effort spent to design one application on
one hardware will not be thrown away when new hardware becomes
available.

In this paper, we describe the PGPG system version 1.0.  In section 2,
we describe the traditional design flow and its problem. In section 3
we describe the basic concept and structure of PGPG.  In section 4, we
show a design of gravitational force pipeline as an example of
pipeline generated by PGPG. Section 5 is for discussion.  Table
\ref{tabab} is a glossary for abbreviations used in the paper.

\begin{table}
\caption{Abbreviation Glossary}
\begin{center}
\begin{tabular}{ll}
\hline
\hline
API & Application Program Interface\\
BEM & Boundary Element Method\\
CAE & Computer Aided Engineering\\
DFT & Discrete Fourier Transform\\
FBCE & FPGA-Based Computing Engine\\
FMM  & Fast Multipole Method\\
FPGA  & Field Programmable Gate Array\\
GRAPE & GRAvity PipE\\
HDL & Hardware Description Language\\
LPM & Library of Parameterized Modules\\
LSI & Large Scale Integration \\
MD  & Molecular Dynamics\\
PGDL & PGPG Description Language [defined in this paper]\\
PGPG  & Pipeline Generator for Programmable GRAPE [defined in this paper]\\
PROGRAPE & PROgrammable GRAPE\\
SLDL & System-Level Design Language\\
SPH  & Smoothed Particle Hydrodynamics\\
VHDL & VHSIC Hardware Description Language \\
VHSIC & Very High Speed Integrated Circuit \\
\hline
\hline
\end{tabular}
\end{center}
\label{tabab}
\end{table}


\section{Traditional Design Flow for FPGA-based Computing Engines}

\label{sectrad}

\begin{figure}
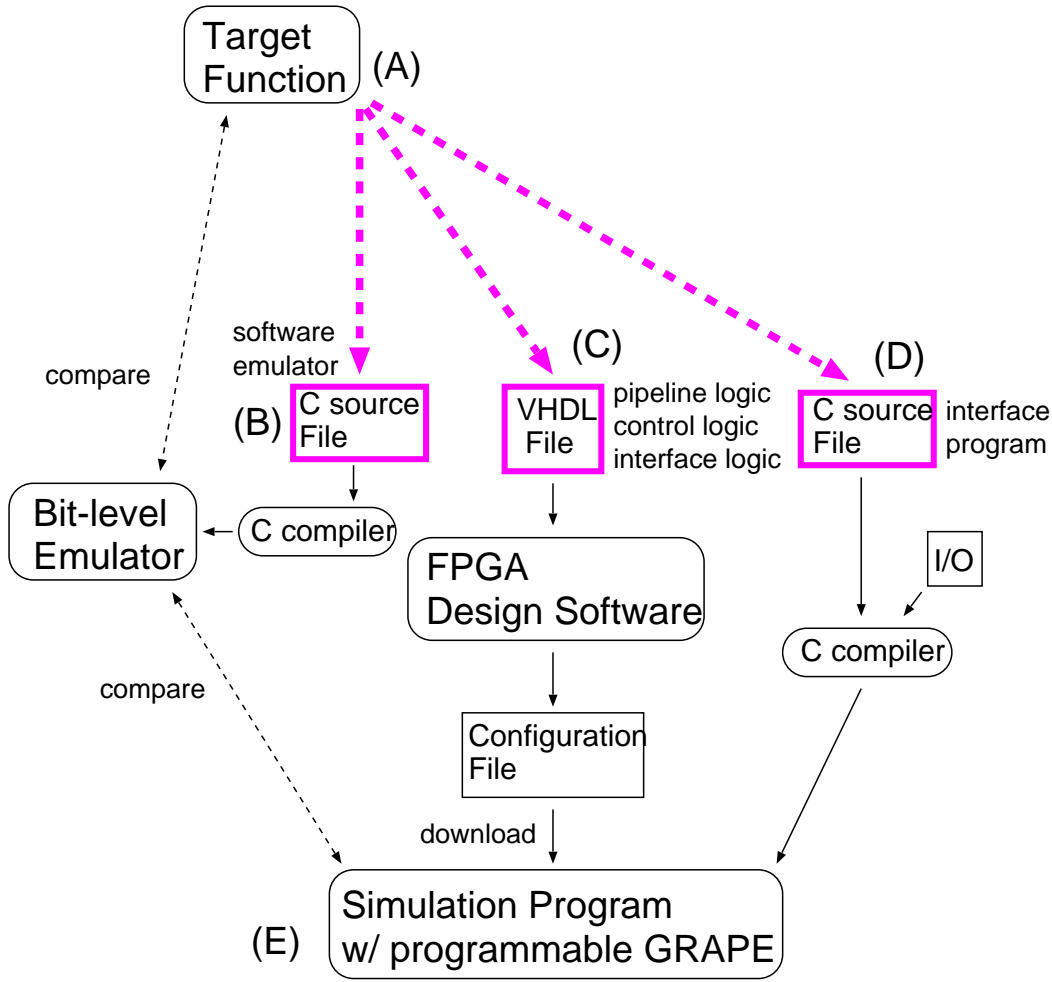

\begin{center}
\leavevmode
\FigureFile(14 cm,12 cm){./figure2.eps}
\end{center}
\caption{Traditional design flow for FPGA-based computing
system. Colored items mean hardware logics and softwares that we have
to develop.}
\label{fig2}
\end{figure}

In the traditional design flow, we design the FPGA-based computing
system in the following five steps.

\begin{itemize}

\setlength{\itemsep}{6pt}

\item[(A)] Target Function Specification:

We specify the target function, namely the function that the pipeline
processor calculates. This includes the specification of the input
data (number format and word length), the dataflow for the calculation
of the function, input and output number format and word length for
each arithmetic operation.

\item[(B)] Bit-Level Software Emulator:

We develop a software emulator which implements the target function
defined in step (A) in software. Using this software emulator, we verify
whether the designed hardware can actually calculate the target
function with required accuracy. In this step, we also define the
application program interface (API). 

\item[(C)]  Hardware Design:

In this step we actually write the source code which implements the
pipeline processor in a hardware-description language (HDL) such as VHDL. In
addition, we design the control logic and host interface logic also
in some HDL. The HDL description is compiled to the configuration data
for the FPGA chips by a design software, usually provided by the
manufacturer of the FPGA device.

\item[(D)] Interface Software:

We develop the software on the host computer which takes care of the
communication to the hardware and data format conversion between the
floating-point data on the host and specialized data format used on
the developed pipeline processor. The developed software should have
the same API as that of the software emulator developed in step (B).

\item[(E)] Finally, we can actually use the pipeline processor with
real application program, by combining the hardware, hardware
configuration data, interface software and application software.

\end{itemize}

Figure \ref{fig2} summarizes these steps.  In these steps, we have to
design, test, and debug a large amount of hardware logic and
softwares. Of course, many of the softwares and hardware designs can
be reused, when we develop different applications. For example, the
design of the floating-point multiplier is rather generic, and can be
used in almost any application. Also, it is possible to buy such
design, or design software which generate floating-point arithmetic
unit with arbitrary word length, from some CAE company. However, just
to understand how to use such a design, one need a deep understanding of
the hardware and HDL used for that particular design. Thus, even
though the reusability significantly reduce the amount of the work
needed for the second and later design for one person, the initial
hurdle remains rather high, for an astrophysicist who never used such
software, or actually the availability of the library make the hurdle
even higher, since a starter need to understand, in addition to the
basics of the hardware design and HDL, the use of such libraries and
particular design software for that library.

The development of the interface software is generally even more
difficult than the design of the hardware, since it requires the
knowledge of how the device driver softwares work in the operating
system of the host computer, and infinite number of small details like
how to integrate the device driver to the operating system, how to
correctly generate the compiler flags to compile the device driver
so that it works with the kernel installed on the host computer etc
etc.

All these works combined make it almost impossible  for an
astrophysicist to even think of implementing the pipeline processor on
an FPGA-based computing engine.


\section{The PGPG system}
\label{secpgpg}

\subsection{Basic Idea of PGPG}
\label{secflow}

\begin{figure}
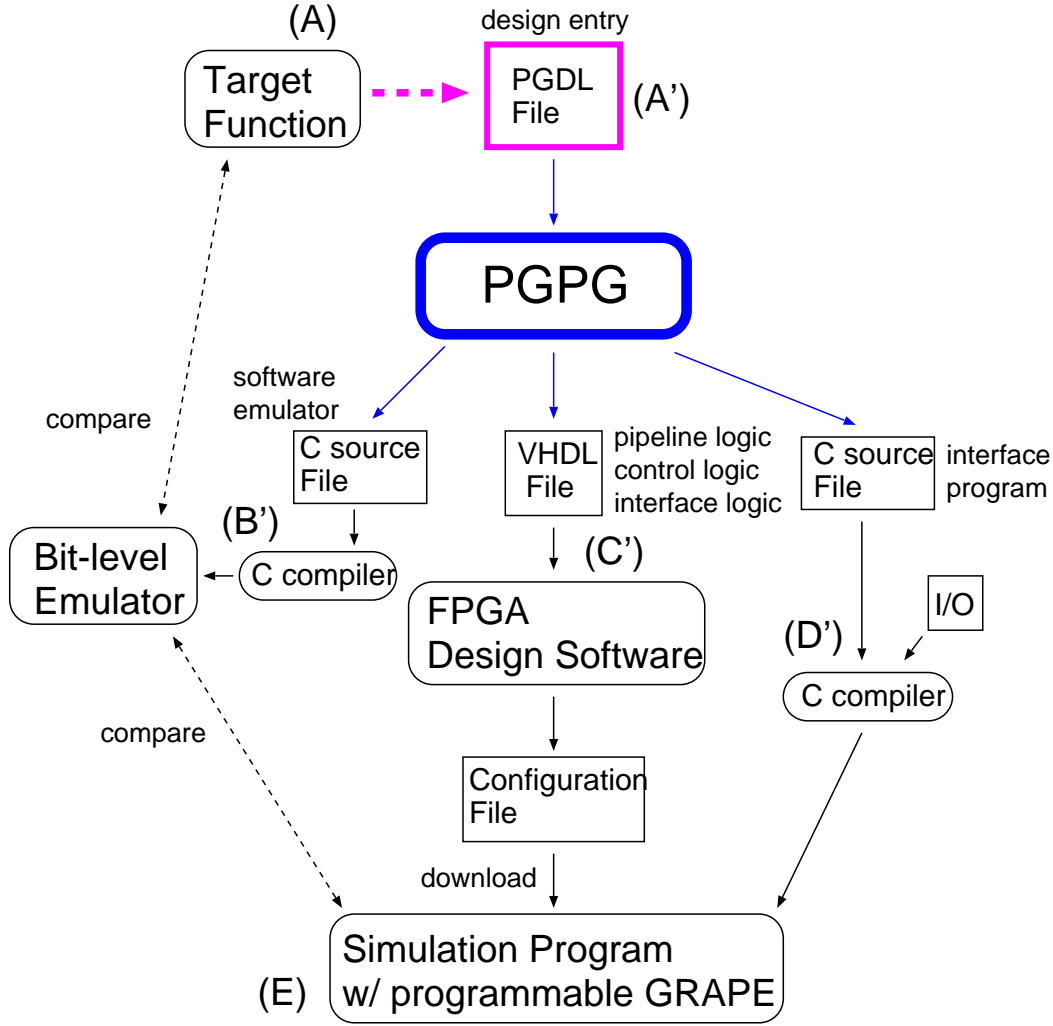

\begin{center}
\leavevmode
\FigureFile(14 cm,12 cm){./figure3.eps}
\end{center}
\caption{Design flow with PGPG. A colored magenta item means the only
file that we have to describe.}
\label{fig3}
\end{figure}

If we inspect  Figure \ref{fig2} again, we can see the fact that {\it
all} softwares and hardware description is derived from the target
function specification in step (A). Thus, it should be possible for a
sufficiently smart software to generate all necessary softwares  and
hardware descriptions from the target function description written in
some high-level language. The basic idea of PGPG is to develop such
a smart software.

Figure \ref{fig3} shows how the design flow changes with PGPG. After
we define the target function, we write it in the high-level specification
language, the PGPG description language (PGDL). The PGPG software
system takes this PGDL description of the pipeline processor as
input, and generates all softwares and hardware descriptions. Thus,
with PGPG, an astrophysicist do not have to write VHDL source code for
the pipeline processor or C source codes for interface library.

In the rest of this section we illustrate how a pipeline processor is
specified in PGDL and how that description is translated to actual
codes.

\subsection{Example Target}

\begin{figure}
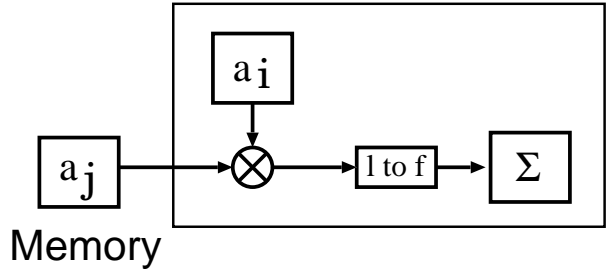

\begin{center}
\leavevmode
\FigureFile(8 cm,6 cm){./figure4.eps}
\end{center}
\caption{Block diagram of the example (artificial) pipeline.}
\label{fig4}
\end{figure}

We consider the following (artificial) example:
\begin{equation}
 f_i = \sum_j^n {a_i a_j}.\qquad (i=1,...,n)
\end{equation}
This function is designed purely to show how the PGDL description and
PGPG software work. Figure \ref{fig4} shows the pipeline
itself. ``Particle'' here is represented by a single scalar value
$a$. The interaction between particles $i$ and $j$ is defined as the
product $a_i a_j$, and we calculate sum over $j$ to obtain the
``force'' on particle $i$. Here, we have the essential ingredients of
the system: particles, their representation, functional form of
interaction.

For the particle data $a_i$ (and $a_j$), we use a logarithmic format,
with 17 bits in total (1 bit for sign, 1 bit for zero or not, 7 bits
for the integer part of logarithm and 8 bits for fractional part). The
base of the logarithm is 2. This logarithmic format has the advantage
that the multiplication becomes addition, so we do not need a
multiplier circuit whose size is $O(m^2)$, where $m$ is the length of
the mantissa. Of course, the addition in logarithm format is more
complex than that in the floating-point format. Thus, the relative
advantage of the log format is not very large. The ``multiplier''
logic itself is generated automatically by PGPG.

In our example target function, we convert the output of multiplier to
fixed-point format, so that we can accumulate it with high
accuracy. This is done by a circuit provided by PGPG. Finally,
converted result is accumulated by a usual fixed-point adder circuit.

The particles with index $j$ is stored in the memory, and new data is
supplied at each clock cycle. The particle with index $i$ is fixed
during one calculation, and is stored in the register within the
pipeline processor.

\begin{figure}
\scriptsize
\begin{verbatim}
#define ascale (pow(2.0,20.0))
#define fscale (1.0/(ascale*ascale))

/NVMP 1;
/NPIPE 2;
/JPSET iaj,aj[],log,17,8,ascale;
/IPSET iai,ai[],log,17,8,ascale;
/FOSET sfij,f[],fix,64,fscale;

pg_log_muldiv(MUL,iaj,iai,aij,17,1);
pg_conv_ltof(aij,fij,17,8,64,1);
pg_fix_accum(fij,sfij,64,64,1);


\end{verbatim}
\caption{An example of design entry file written in PGDL}
\label{fig5}
\end{figure}

Figure \ref{fig5} shows the PGDL description of this target function.
The first two lines define formulae used for the data format conversion
between the internal data format (logarithmic for $a_i$ and
fixed-point for $f_i$). These are actually used in the next block,
which defines the interface etc.  The next block (lines starting with
``/'') defines the register and memory layout, which also determines
API.  The final part describes the target function itself. It has
C-like appearance, but actually defines the hardware modules and their
interconnection.  In the next subsection we describe the PGPG language
in more detail.

\subsection{The PGDL Language}

In this section, we give a minimal description of the PGDL language. A
full description is available in {\tt http://progrape.jp}.

\subsubsection{PGDL Target Hardware Model}

\begin{figure}
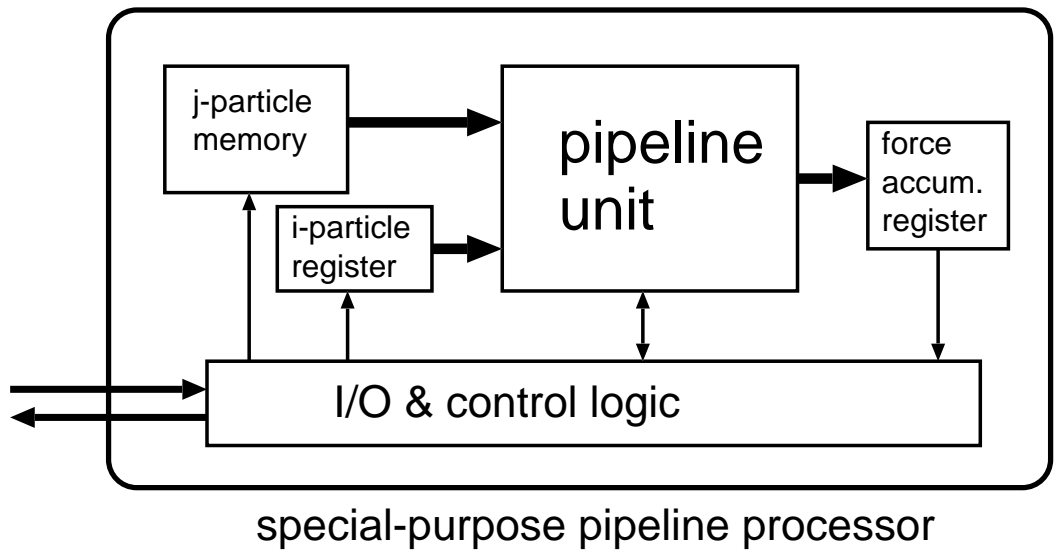

\begin{center}
\leavevmode
\FigureFile(14 cm,12 cm){./figure5.eps}
\end{center}
\caption{Block diagram of the special-purpose processor
generated from a PGDL program}
\label{fig5a}
\end{figure}

Figure \ref{fig5a} gives the structure of the special-purpose
processor generated from a PGDL ``program''. It consist of the control
logic, I/O logic, program-specified registers and a memory unit, and
the pipeline unit. Program-specified registers are either input
registers, which we call $i$-particle registers, or registers which
accumulate the calculated interaction, which we call
force-accumulation registers. We call memory unit $j$-particle memory.

In figure \ref{fig5}, the pipeline processor is specified by list of
modules (lines with {\tt pg\_...}). Registers and memories are
specified by lines with {\tt /IPSET} ($i$-particle register), {\tt
/JPSET} ($j$-particle memory), and {\tt /FOSET} (force-accumulation
register).

This hardware model is general and flexible enough to express any
special-purpose computer which calculates the function of the form of
equation (\ref{eq:PGPGbasic}). We use the analogy of particles and
forces, but the actual data in the $i$-particle register or
$j$-particle memory need not represent physical particles, and
``force'' can be something completely different. For example, this
hardware model can be used to describe Discrete Fourier Transform
(DFT) or other types of convolution operations.

\subsubsection{PGDL API Model}

Currently, one PGDL program generates one function prototype, {\tt
void force(...)}, with list of arguments. The
list of arguments consists of the data to be stored to $i$-particle
registers, that to be stored to $j$-particle memories, and the data to
be returned (the content of force accumulation registers). For the
body of the function, both the emulator function and actual driver and
data conversion function are generated, and the user can use either
one by linking appropriate object file, without changing the source
file.  In figure \ref{fig5}, second arguments of {\tt /JPSET}, {\tt
/IPSET} and {\tt /FOSET} lines determine the name of the arguments
which corresponds to the specified register or memory element.

\subsubsection{PGDL Program Structure}

A PGDL program consists of the following sections:

\begin{enumerate}
\item macro declaration
\item generic declaration
\item interface declaration
\item pipeline description
\end{enumerate}

The macro declaration, which is the first two lines of code in figure
\ref{fig5}, is processed by C preprocessor (cpp) and used just for
convenience to allow the same expressions which appear in multiple
places to be defined only once.

The generic declaration in figure \ref{fig5} are two
lines:

\begin{verbatim}
/NVMP 1;
/NPIPE 2;
\end{verbatim}

The first line determines the degree of the virtual multiple pipeline
(Makino et al. 1997). The second one is the number of physical
pipelines implemented to the current design. Thus, we can change the
physical number of pipelines by just change this parameter, and the
application program can make use of the parallel pipeline without any
need to change the user code.

The interface declaration is the following part:

\begin{verbatim}
/JPSET iaj,aj[],log,17,8,ascale;
/IPSET iai,ai[],log,17,8,ascale;
/FOSET sfij,f[],fix,64,fscale;
\end{verbatim}

The first argument is the name used for the registers and memories in
the pipeline description, and the second one is the name used for
API. the remaining arguments specifies the number formats. In this
example, both $a_i$ and $a_j$ are in the logarithmic format, with 17
bits of the total word length and 8 bits of mantissa. 

Finally, the pipeline description is the following part:

\begin{verbatim}
pg_log_muldiv(MUL,iaj,iai,aij,17,1);
pg_conv_ltof(aij,fij,17,8,64,1);
pg_fix_accum(fij,sfij,64,64,1);
\end{verbatim}

Here, {\tt pg\_log\_muldiv} generates one multiplier in the
logarithmic format, which takes two inputs, {\tt iaj} and {\tt iai}, and
calculates one output result, {\tt aij}. The rest of arguments, {\tt
17} and {\tt 1} indicate the bit length and number of pipeline stages,
respectively.  The inputs are taken from the $j$-particle memory and
the $i$-particle register with the corresponding names, and the output becomes
the input to the next module {\tt pg\_conv\_ltof}. This module
converts the logarithmic format to fixed-point formant.  Finally,
module {\tt pg\_fix\_accum} accumulates the result, and the value of
this accumulator, {\tt sfij} is accessible from the application
program with name {\tt f}, as specified in {\tt /FOSET} declaration.

\subsubsection{PGDL Arithmetic Modules}

The present version of PGDL supports the following two number format:
(a) fixed-point format and (b) logarithmic format. For the fixed-point
format PGDL supports addition (and accumulation as well), subtraction,
and conversion to the logarithmic format. For the logarithmic format,
multiplication, division, power functions (with rational powers), and
conversion to fixed-point format are supported. Appendix 1 gives more
detailed discussion of the PGDL language elements.


\section{A Real Example: Gravitational Force Pipeline}
\label{secgrav}

In this section, we discuss the implementation of the gravitational
force calculation in PGDL in detail. We chose the gravitational force
calculation as the example target, because we can compare the
performance and size of PGDL-generate design with hand-coded ones such
as the pipeline design of GRAPE-5. 

The pipeline to be designed calculates the gravitational force on
particle $i$:
\begin{equation}
\mathbf{ a}_i = \sum_j {m_j \mathbf{ r}_{ij} \over (r_{ij}^2 + \varepsilon^2)^{3/2}}
\end{equation}
where $\mathbf{ a}_i$ is the gravitational acceleration of particle $i$,
$\mathbf{ r}_i$ and $m_i$ are the position and mass of particle $i$,
$\mathbf{ r}_{ij} = \mathbf{ r}_{j} - \mathbf{ r}_{i}$, and $\varepsilon$ is a
softening parameter. Here we design the pipeline essentially the same
as that of GRAPE-3 and GRAPE-5, to compare the performance and size.

\subsection{The PGDL Pipeline Design Description}

\begin{figure}
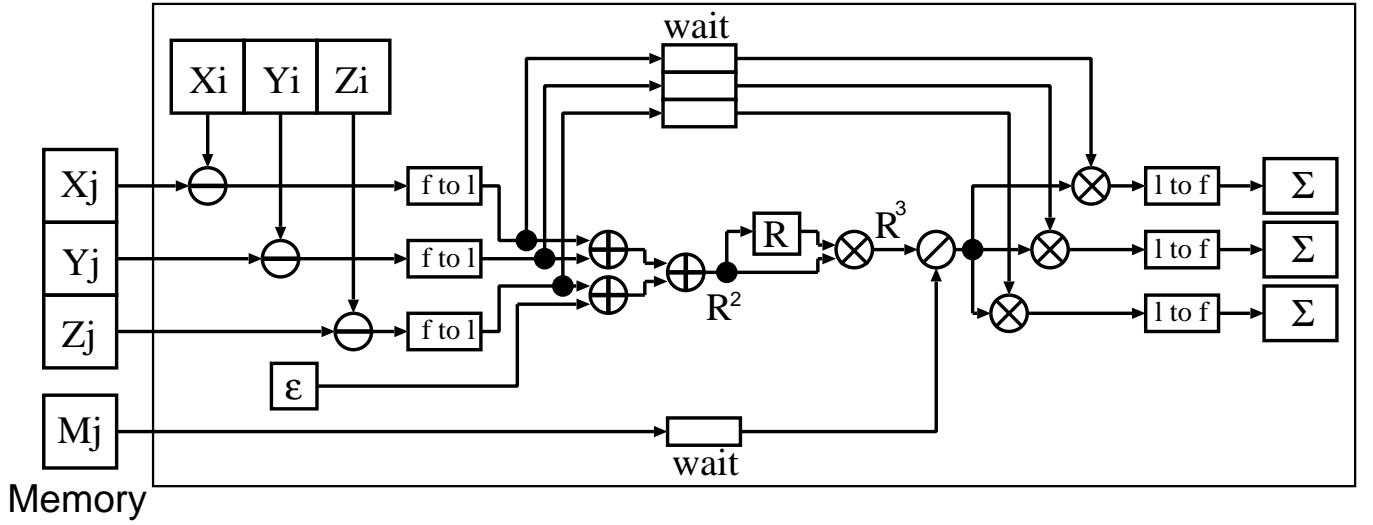

\begin{center}
\leavevmode
\FigureFile(18 cm,12 cm){./figure6.eps}
\end{center}
\caption{Block diagram of the pipeline for gravitational force}
\label{fig7}
\end{figure}

Figure \ref{fig7} shows the block diagram of the gravitational force
pipeline. Position data for both $i$-particle and $j$-particle are
in the fixed-point format, while $m_j$ is in the logarithmic
format. After subtraction, $\mathbf{x}_j-\mathbf{x}_i$, the results
are converted to the logarithmic format, and all calculations until
the final accumulation are done in this logarithmic format.

Figure
\ref{fig8} shows the PGDL program file for the gravitational force
pipeline in figure \ref{fig7}.
One can see that each {\tt pg} module in figure \ref{fig8} directly
corresponds to arithmetic units in figure \ref{fig7}. Actually, the
PGDL description is more compact, since it allows implicit array
operation, as in the case of the first line:
\begin{verbatim}
pg_fix_addsub(SUB,xi,xj,xij,NPOS,1);
\end{verbatim}
Here, {\it three} modules are generated automatically, because both
{\tt xi} and {\tt xj} are declared as the array of size 3 in the
interface declaration. Also, there is no need to explicitly specify
the wait (pipeline delay) modules, since the PGDL compiler inserts the
necessary delay element automatically.

\begin{figure}
\scriptsize
\begin{verbatim}
#define xscale (pow(2.0,32.0)/64.0)
#define mscale (pow(2.0,60.0)/(1.0/1024.0))
#define escale (xscale*xscale)
#define fscale (-xscale*xscale/mscale)
#define NPOS 32
#define NLOG 17
#define NMAN 8
#define NFOR 57
#define NACC 64

/NVMP 2;
/NPIPE 2;
/JPSET xj[3],x[][],ufix,NPOS,xscale;
/JPSET mj,m[],log,NLOG,NMAN,mscale
/IPSET xi[3],x[][],ufix,NPOS,xscale;
/IPSET ieps2,eps2,log,NLOG,NMAN,escale;
/FOSET sx[3],a[][],fix,NACC,fscale;

pg_fix_addsub(SUB,xi,xj,xij,NPOS,1);
pg_conv_ftol(xij,dx,NPOS,NLOG,NMAN,4);
pg_log_shift(1,dx,x2,NLOG);
pg_log_unsigned_add(x2[0],x2[1],x2y2,NLOG,NMAN,3);
pg_log_unsigned_add(x2[2],ieps2,z2e2,NLOG,NMAN,3);
pg_log_unsigned_add(x2y2,z2e2,r2,NLOG,NMAN,3);
pg_log_shift(-1,r2,r1,NLOG);
pg_log_muldiv(MUL,r2,r1,r3,NLOG,1);
pg_log_muldiv(DIV,mj,r3,mf,NLOG,1);
pg_log_muldiv(MUL,mf,dx,fx,NLOG,1);
pg_conv_ltof(fx,ffx,NLOG,NMAN,NFOR,2);
pg_fix_accum(ffx,sx,NFOR,NACC,2);

\end{verbatim}
\caption{A sample design entry file for the gravitational force
pipeline written in PGDL}
\label{fig8}
\end{figure}

The top-level interface function to the application program generated
from this PGDL program has the following form:

\begin{verbatim}
void force(double x[][3], double m[], double eps2, double a[][3], int n);
\end{verbatim}
In this example, positions of $i$ particles and $j$ particles are
passed as a single array {\tt x}, since the same name is used in {\tt
/IPSET} and {\tt /JPSET}. Thus, with this interface it is only
possible to calculate the self-gravity of an $N$-body system.


\subsection{Performance of Generated Pipeline}

Here we report the performance of the PGDL-generated gravitational
force calculation pipeline, and compare that with that of GRAPE-3 and
GRAPE-5.  We summarize internal number expressions for the pipeline
designs in table \ref{tab2}.

\begin{table}
\caption{Model}
\begin{center}
\begin{tabular}{lccc}
\hline
\hline
Model& Position & Internal(mantissa)  & Accumulation\\
\hline
G3  & 20bit fixed  & 14(5)bit log & 56bit fixed\\
G5  & 32bit fixed  & 17(8)bit log & 64bit fixed\\
G5+ & 32bit fixed  & 20(11)bit log & 64bit fixed\\
\hline
\hline
\end{tabular}
\end{center}
\label{tab2}
\end{table}

Table \ref{tab3} shows the size and performance of the generated
pipeline (model G5) for several implementations with different number
of pipeline stages, for two different kinds of the Altera device. The
pipeline designs with different number of pipeline stages can be
easily obtained by a small modification of the design entry file in
PGPG.  The size and maximum operation speeds are those reported by
Altera's design software, Quartus II(ver 3.0). The speed grade of
these devices are (-2) for APEX20k and (-5) for Stratix (fastest
available at the time of the writing).

\begin{table}
\caption{Performance of the generated pipeline (model G5)}
\begin{center}
\begin{tabular}{ccc|ccc}
\hline
\hline
 & APEX20k & & & Stratix & \\
stage & size(LE) & $f_{\rm max}$(MHz) & stage & size(LE) & $f_{\rm max}$(MHz) \\
\hline
14 & 2735 & 58.92 & 17 & 2499 & 133.30 \\
16 & 2928 & 60.97 & 19 & 2655 & 137.51 \\
17 & 2925 & 73.78 & 21 & 2849 & 142.29 \\
20 & 3074 & 74.65 & 23 & 2927 & 135.78 \\
21 & 3064 & 80.33 & 24 & 2864 & 142.88 \\
\hline
\hline
\end{tabular}
\end{center}
\label{tab3}
\end{table}

\begin{table}
\caption{Performance of generated pipelines (Stratix)}
\begin{center}
\begin{tabular}{lcccc}
\hline
\hline
Model & $f_{\rm max}$ & Size & Memory & Stage \\
 & (MHz) & (LE) & (bit) & \\
\hline
G5 & 181.82 & 3021 & 41k & 30 \\
G3 & 191.64 & 2369 & 21k & 26 \\
G5+ & 142.27 & 5082 & 402k & 35 \\
\hline
\hline
\end{tabular}
\end{center}
\label{tab4}
\end{table}

Table \ref{tab4} shows the size and performance of pipelines with
different accuracy (G3, G5 and G5+). One can see that both the performance
penalty and size increase due to increased accuracy of G5+, compared
to G3 or G5, are fairly modest.

With currently available FPGA (Stratix EP1S20), we can fit 5 G5
pipelines running at 180 MHz into one chip. The original GRAPE-5
pipeline chip, which was made 7 years ago, had two pipelines operating
at 80 MHz clock. Thus, FPGA implementation of GRAPE-5 has about 5
times more speed than the original custom-chip implementation. The
peak speed of one FPGA chip for GRAPE-5 pipeline is 34.2 Gflops.  Of
course, this large improvement over GRAPE-5 is due primarily to the
advance in the semiconductor technology in the 7 years (from $0.5{\rm
\mu m}$ to 130 nm), but clearly indicates that FPGA-based computing
engine does offer very good performance, and that PGDL provides
a practical tool to implement special-purpose computers on FPGA-based
computing engines.


\section{Discussion}

\subsection{Comparison with Other Design Methodology}

In other area, such as digital signal processing, there exist many
code generators that generates HDL code from a simple description.
For example, a commercial package (MATLAB) generates HDL code for
fixed-point filter designed using itself.

Recently, a design methodology called the system-level design has
become popular.  In the system-level design, the function and
architecture of LSI or FPGA are described using C/C++ languages or
subset of them.  These languages are called the System-Level
Description Language (SLDL). Using SLDL, programmers can verify
functionality and performance at the early stage of the development.
The design is divided into software part and hardware part.  The
hardware part will be synthesized to register-transfer design by a
SLDL design software. The SpecC, System-C, or Handel-C are the
well-known SLDLs commercially available. There are also a number of
research projects to design hardware using C++/Java languages
(e.g. Hutchings 1999, Mencer 2002, Tsoi 2004).

The goal of these SLDL is to describe hardware logic without using hardware
description language, such as VHDL or Verilog HDL. Therefore, even if
we use SLDL, we still need a detailed description of the hardware in
other language like C or C++.  If we consider in the traditional flow
(Figure \ref{fig2}), we can save step (C) and (D) using SLDL, while
step (B) is still required.  Using PGPG, we can replace all of steps
by writing a short high-level hardware description.

\subsection{Planned and Ongoing Improvement of PGPG and PGDL}

In this paper, we described basic concept and functions of PGPG.  For
those who are more interested, we put a CGI program of the current
version of PGPG at a web site ({\tt http://progrape.jp}).  The CGI
program generates VHDL code, user interface code, and emulator code from
a PGDL description. 

Although the current version of PGPG is successful in designing the
pipeline for the gravitational force, its functionality is rather
limited. We are currently developing the next version of PGPG that
supports more functionality and multiple hardwares.  For the next
version, we plan to add the modules needed to design a pipeline for
the SPH simulation and Boundary Element Method (BEM). BEM is one of
methods to solve numerically the boundary value problem of partial
differential equation (Brebbia 1978). The floating-point format with
longer mantissa is needed for these applications. We are now further
developing the support of the floating-point arithmetic module for
PGPG. We also plan to support Xilinx FPGA chips as well as Altera
chips. As the target board for Xilinx device, we use Bioler3/HORN-5
board, developed by Chiba University and RIKEN (Ito et al 2004).

\bigskip

This research was partially supported by the Grants-in-Aid by the
Japan Society for the Promotion of Science (14740127) and by the
Ministry of Education, Science, Sports, and Culture of Japan
(16684002).

\appendix
\section{Description of PGDL Declarations and Available Modules}

\begin{table}
\caption{PGPG version 1.0 feature}
\begin{center}
\begin{tabular}{lll}
\hline
Module & {\tt pg\_fix\_addsub} &  fixed point format adder/subtracter\\
 & {\tt pg\_fix\_accum} &  fixed point format accumulator\\
 & {\tt pg\_log\_unsigned\_add} &  unsigned logarithmic format adder\\
 & {\tt pg\_log\_muldiv} &  logarithmic format multiplier/divider\\
 & {\tt pg\_log\_shift} &  logarithmic format shifter\\
 & {\tt pg\_conv\_ftol} &  converter from fixed point format to logarithmic format\\
 & {\tt pg\_conv\_ltof} &  converter from logarithmic format to fixed point format\\
\hline
Definition & {\tt /NPIPE} &  number of pipeline \\
 & {\tt /NVMP} &  number of virtual multiple pipeline \\
 & {\tt /JPSET} &  memory unit setting\\
 & {\tt /IPSET} &  input register setting\\
 & {\tt /FOSET} &  output register setting\\
\hline
Device support& & Altera's FPGA\\
Hardware support & & PROGRAPE-2 \\
Other & & Options for look-up table\\
\hline
\end{tabular}
\end{center}
\label{tab1}
\end{table}

Table \ref{tab1} shows the features of the current version of PGPG.
The specification of version 1.0 is determined so that a pipeline for
gravitational force, shown in section \ref{secgrav}, can be constructed as
the first step.

PGPG version 1.0 supports nine parametrized modules as shown in Table
\ref{tab1}. The bit length and the number of pipeline stage for each
module can be changed by the arguments. For example, the arguments of
the fixed point format adder/subtracter {\tt pg\_fix\_addsub(SUB,xi,xj,xij,32,1)}
indicate an operation flag(adder or subtracter), the first input, the
second input, output, bit length, and number of pipeline stages,
respectively, from the first to sixth argument.

Modules {\tt pg\_fix\_addsub} and {\tt pg\_fix\_accum} are fixed point
format adder/subtracter and sign-magnitude accumulator, respectively.
Modules {\tt pg\_log\_muldiv} and {\tt pg\_log\_unsigned\_add} are
logarithmic format multiplier/divider and unsigned adder,
respectively. In the logarithmic format, a positive, non-zero real
number $x$ is represented by its base-2 logarithm $y$ as
$x=2^{y}$. The logarithmic format has been adapted for the
gravitational pipeline because it has larger dynamics length for the
same word length and operation such as multiplication and square root
are easier to implement than in the usual floating-point format. For
more details of the logarithmic format, see GRAPE-5 paper (Kawai et
al. 2000).  Module {\tt pg\_log\_shift} is a logarithmic format
shifter. Shift operations in the logarithmic format express square
(left shift) and squared root (right shift). Module {\tt
pg\_conv\_ftol} is a converter from the fixed point format to the
logarithmic format, and {\tt pg\_conv\_ltof} is a converter from the
logarithmic format to the fixed point format.  In PGPG version 1.0,
these modules are described partly using the Altera's LPM.  
A gap of delay timing is synchronized automatically by the PGPG.

Addition to the parametrized modules, five definitions are defined in
PGPG version 1.0.  Definitions {\tt /NPIPE} and {\tt /NVMP} define the
numbers of (real) pipeline and virtual multiple pipeline (Makino et
al. 1997), respectively. Definition {\tt /JPSET} defines a setting for
the memory unit. Definitions {\tt /IPSET} and {\tt /FOSET} define
settings for the input and output registers in the interaction
pipeline, respectively.

\section{Details of The Generated Gravitational Force Pipeline}

In this appendix, we show a part of the code generated by PGPG from PGDL
description of the force calculation pipeline.  More complete code is
obtained by a CGI program of the current version of PGPG in a website
({\tt http://progrape.jp}). 

\subsection{VHDL Code}

PGPG generates description files of the designed hardware logic in
VHDL.  The hardware logic includes the pipeline logic itself and its
peripheral logic. Figures \ref{fig11} and \ref{fig12} show a part of
the VHDL source files generated by PGPG (the total length is about
2800 lines).  The design software provided by the FPGA manufacture
creates configuration data of FPGA from the generated sources [Step
(C') in figure \ref{fig3}]. The configuration data are downloaded into
the programmable GRAPE hardware using the interface program also
generated by PGPG.

\begin{figure}
\scriptsize
\begin{verbatim}
library ieee;                                                       
use ieee.std_logic_1164.all;                                        
use ieee.std_logic_unsigned.all;                                    
                                                                    
entity pipe is                                                      
  generic(JDATA_WIDTH : integer :=72) ;                             
port(p_jdata : in std_logic_vector(JDATA_WIDTH-1 downto 0);         
     p_run : in std_logic;                                          
     p_we :  in std_logic;                                          
     p_adri : in std_logic_vector(3 downto 0);                      
     p_adrivp : in std_logic_vector(3 downto 0);                    
     p_datai : in std_logic_vector(31 downto 0);                    
     p_adro : in std_logic_vector(3 downto 0);                      
     p_adrovp : in std_logic_vector(3 downto 0);                    
     p_datao : out std_logic_vector(31 downto 0);                   
     p_runret : out std_logic;                                      
     rst,pclk : in std_logic);                                      
end pipe;                                                           
                                                                    
architecture std of pipe is                                         

begin                                                               
                                                                    
  process(pclk) begin
    if(pclk'event and pclk='1') then
      jdata1 <= p_jdata;            
    end if;
  end process;
              
  process(pclk) begin
    if(pclk'event and pclk='1') then
      if(vmp_phase = "0000") then
        xj(31 downto 0) <= jdata1(31 downto 0);
        yj(31 downto 0) <= jdata1(63 downto 32);
        zj(31 downto 0) <= p_jdata(31 downto 0);
        mj(16 downto 0) <= p_jdata(48 downto 32);
      end if;
    end if;
  end process;
     .
     .
     .

  u0: pg_fix_sub_32_1  port map (x=>xi,y=>xj,z=>xij,clk=>pclk);
  u1: pg_fix_sub_32_1  port map (x=>yi,y=>yj,z=>yij,clk=>pclk);
  u2: pg_fix_sub_32_1  port map (x=>zi,y=>zj,z=>zij,clk=>pclk);
  u3: pg_conv_ftol_32_17_8_4 port map (fixdata=>xij,logdata=>dx,clk=>pclk);
  u4: pg_conv_ftol_32_17_8_4 port map (fixdata=>yij,logdata=>dy,clk=>pclk);
  u5: pg_conv_ftol_32_17_8_4 port map (fixdata=>zij,logdata=>dz,clk=>pclk);
     .
     .
     .

end std;                                                            
\end{verbatim}
\caption{A part of the source files in VHDL (part 1) for the pipeline
logic generated from the design entry file shown in Figure
\ref{fig8}. Component and signal declaration sentences are omitted.}
\label{fig11}
\end{figure}

\begin{figure}
\scriptsize
\begin{verbatim}
library ieee;                                                      
use ieee.std_logic_1164.all;                                       
                                                                   
entity pg_conv_ftol_32_17_8_4 is         
  port(fixdata : in std_logic_vector(31 downto 0);      
       logdata : out std_logic_vector(16 downto 0);     
       clk : in std_logic);                                        
end pg_conv_ftol_32_17_8_4;              
                                                                   
architecture rtl of pg_conv_ftol_32_17_8_4 is 
                                                                   
begin                                                              
                                                                   
  d1 <=  NOT fixdata(30 downto 0);                     
  one <= "0000000000000000000000000000001";                                              
  u1: lpm_add_sub generic map (LPM_WIDTH=>31,LPM_DIRECTION=>"ADD")
                  port map(result=>d2,dataa=>d1,datab=>one);     
  d0 <= fixdata(30 downto 0);                        
  sign0 <= fixdata(31);                              
                                                                 
  with sign0 select                                              
    d3 <= d0 when '0',                                           
    d2 when others;                                              
                                                                 
  process(clk) begin                                             
    if(clk'event and clk='1') then                               
      d3r <= d3;                                                 
      sign1 <= sign0;                                            
    end if;                                                      
  end process;                                                   
                                                                 
  u2: penc_31_5 port map (a=>d3r,c=>c1);   
  with d3r select                                                
    nz0 <= '0' when "0000000000000000000000000000000",                                  
           '1' when others;                                      
     .
     .
     .
                                                                 
end rtl;                                                         
\end{verbatim}
\caption{A part of the source file in VHDL (part 2) for the pipeline logic 
generated from the design entry file shown in Figure \ref{fig8} (part
2). Component and signal declaration sentences are omitted.} 
\label{fig12}
\end{figure}

\subsection{Interface Functions}

The interface software on the host computer to the programmable GRAPE
hardware is composed by the C complier of the sources generated by
PGPG [Step(D') in figure \ref{fig3}]. Figure \ref{fig13} shows a part
of source files in C generated by PGPG (the total length is about 150
lines).  We run the application program linked with the interface
software [Step(E)].

\begin{figure}
\scriptsize
\begin{verbatim}
#include <stdio.h>
#include <math.h>

void force(double x[][3], double m[], double eps2, double a[][3], int n)
{
  npipe = 4;
  pgpgi_initial();
  pgpgi_setxj(n,x,m);

  for(i=0;i<n;i+=npipe){
    if((i+npipe)>n){
      nn = n - i;
    }else{
      nn = npipe;
    }

    pgpgi_setxi(i,nn,x,eps2);
    pgpgi_run(n);
    pgpgi_getforce(i,nn,a);
  }

}

void pgpgi_setxj(int n, double x[][3], double m[])
{
  devid = 0;
  for(j=0;j<n;j++){
    xj = ((unsigned int) (x[j][0] * (pow(2.0,32.0)/(64.0)) + 0.5)) & 0xffffffff;
    yj = ((unsigned int) (x[j][1] * (pow(2.0,32.0)/(64.0)) + 0.5)) & 0xffffffff;
    zj = ((unsigned int) (x[j][2] * (pow(2.0,32.0)/(64.0)) + 0.5)) & 0xffffffff;
    if(m[j] == 0.0){
      mj = 0;
    }else if(m[j] > 0.0){
      mj = (((int)(pow(2.0,8.0)*log(m[j]*(pow(2.0,60.0)/(1.0/1024.0)))/log(2.0))) & 0x7fff) | 0x8000;
    }else{
      mj = (((int)(pow(2.0,8.0)*log(-m[j]*(pow(2.0,60.0)/(1.0/1024.0)))/log(2.0))) & 0x7fff) | 0x18000;
    }

    nword = 4;
    jpdata[0] = 0xffc00;
    jpdata[1] = 2*j+1;
    jpdata[2] = 0x0 | ((0xffffffff & xj) << 0) ;
    jpdata[3] = 0x0 | ((0xffffffff & yj) << 0) ;
    g6_set_jpdata(devid,nword,jpdata);

    jpdata[1] = 2*j+0;
    jpdata[2] = 0x0 | ((0xffffffff & zj) << 0) ;
    jpdata[3] = 0x0 | ((0x1ffff & mj) << 0) ;
    g6_set_jpdata(devid,nword,jpdata);
  }
}

\end{verbatim}
\caption{A part of source file in C for the interface program generated
from the design entry file shown in Figure \ref{fig8}.  Variable
declaration sentences are omitted. }
\label{fig13}
\end{figure}

\subsection{Emulator Code}

Figures \ref{fig9} and \ref{fig10} show a part of the source
files generated by PGPG (the total length is 580 lines).

\begin{figure}
\scriptsize
\begin{verbatim}
#include <stdio.h>
#include <math.h>

void force(double x[][3], double m[], double eps2, double a[][3], int n)
{                                                                   
  for(i=0;i<n;i++){                                               
    xi = ((unsigned int) (x[i][0] * (pow(2.0,32.0)/64.0) + 0.5)) & 0xffffffff;
    yi = ((unsigned int) (x[i][1] * (pow(2.0,32.0)/64.0) + 0.5)) & 0xffffffff;
    zi = ((unsigned int) (x[i][2] * (pow(2.0,32.0)/64.0) + 0.5)) & 0xffffffff;
    if(eps2 == 0.0){                                         
      ieps2 = 0;                                                
    }else if(eps2 > 0.0){                                    
      ieps2 = (((int)(pow(2.0,8.0)*log(eps2*((pow(2.0,32.0)/64.0)*(pow(2.0,32.0)/64.0)))/log(2.0))) & 0x7fff) | 0x8000;
    }else{                                                 
      ieps2 = (((int)(pow(2.0,8.0)*log(-eps2*((pow(2.0,32.0)/64.0)*(pow(2.0,32.0)/64.0)))/log(2.0))) & 0x7fff) | 0x18000;
    }                                                      
    sx = 0;
    sy = 0;
    sz = 0;
                                                                    
    for(j=0;j<n;j++){                                             
      xj = ((unsigned int) (x[j][0] * (pow(2.0,32.0)/64.0) + 0.5)) & 0xffffffff;
      yj = ((unsigned int) (x[j][1] * (pow(2.0,32.0)/64.0) + 0.5)) & 0xffffffff;
      zj = ((unsigned int) (x[j][2] * (pow(2.0,32.0)/64.0) + 0.5)) & 0xffffffff;
      if(m[j] == 0.0){                                         
        mj = 0;                                                
      }else if(m[j] > 0.0){                                    
        mj = (((int)(pow(2.0,8.0)*log(m[j]*(pow(2.0,60.0)/(1.0/1024.0)))/log(2.0))) & 0x7fff) | 0x8000;
      }else{                                                 
        mj = (((int)(pow(2.0,8.0)*log(-m[j]*(pow(2.0,60.0)/(1.0/1024.0)))/log(2.0))) & 0x7fff) | 0x18000;
      }                                                      
                                                                    
      pg_fix_sub_32(xi,xj,&xij);
      pg_fix_sub_32(yi,yj,&yij);
      pg_fix_sub_32(zi,zj,&zij);
      pg_conv_ftol_fix32_log17_man8(xij,&dx);
      pg_conv_ftol_fix32_log17_man8(yij,&dy);
      pg_conv_ftol_fix32_log17_man8(zij,&dz);
        .
        .
        .
      pg_fix_accum_f57_s64(ffx,&sx);
      pg_fix_accum_f57_s64(ffy,&sy);
      pg_fix_accum_f57_s64(ffz,&sz);
    }                                                                 
    a[i][0] = ((double)(sx<<0))*(-(pow(2.0,32.0)/64.0)*(pow(2.0,32.0)/64.0)/(pow(2.0,60.0)/(1.0/1024.0)))/pow(2.0,0.0);
    a[i][1] = ((double)(sy<<0))*(-(pow(2.0,32.0)/64.0)*(pow(2.0,32.0)/64.0)/(pow(2.0,60.0)/(1.0/1024.0)))/pow(2.0,0.0);
    a[i][2] = ((double)(sz<<0))*(-(pow(2.0,32.0)/64.0)*(pow(2.0,32.0)/64.0)/(pow(2.0,60.0)/(1.0/1024.0)))/pow(2.0,0.0);
  }                                                                   
}                                                                   
\end{verbatim}
\caption{The source file (for top architecture) in C for bit-level emulator 
generated from the design entry file shown in Figure \ref{fig8}. 
Variable declaration sentences are omitted.}
\label{fig9}
\end{figure}

\begin{figure}
\scriptsize
\begin{verbatim}
#include<stdio.h>
#include<math.h>
void pg_conv_ftol_fix32_log17_man8(int fixdata, int* logdata){

  /* SIGN BIT */
  fixdata_msb = 0x1&(fixdata >>31);
  logdata_sign = fixdata_msb;

  /* ABSOLUTE */
  fixdata_body    = 0x7FFFFFFF & fixdata;
  {
    int inv_fixdata_body=0;
    inv_fixdata_body = 0x7FFFFFFF ^ fixdata_body;
    if(fixdata_msb == 0x1){
      abs = 0x7FFFFFFF & (inv_fixdata_body + 1);
    }else{
      abs = fixdata_body;
    }
  }
  abs_decimal = 0x3FFFFFFF& abs;

  /* GENERATE NON-ZERO BIT (ALL BIT OR) */
  if(abs != 0x0){ logdata_nonzero = 0x1; }else{ logdata_nonzero=0x0; }

  { /* PRIORITY ENCODER */
    int i;
    int count=0;
    for(i=31;i >=0;i--){
        int buf;
        buf = 0x1 & (abs >>i);
        if(buf == 0x1){ count = i; break;}
        count = i;
    }
    penc_out=count;
  }
  penc_out = 0x1F & penc_out; /* 5-bit */

    .
    .
    .

}
\end{verbatim}
\caption{A part of the source file (for modules) in C for the bit-level emulator
generated from the design entry file shown in Figure \ref{fig8}. 
Variable declaration sentences are omitted.}
\label{fig10}
\end{figure}

\end{document}